# Ion-scale Solitary Structures in the Solar Wind Observed by Solar Orbiter and Parker Solar Probe

Yufei Yang (杨宇菲)[1], Timothy S. Horbury[1], Domenico Trotta[1,2], Lorenzo Matteini[1], Joseph H. Wang[1], Andrey Fedorov[3], Philippe Louarn[3], Stuart D. Bale[4,5], Marc Pulupa[5], Davin E. Larson[5], Roberto Livi[5], Michael L. Stevens[6], Milan Maksimovic[7], Yuri V. Khotyaintsev[8,9], and Andrea Larosa[10]

[1] Department of Physics, Imperial College London, London, SW7 2BW, UK
[2] European Space Agency, European Space Astronomy Centre, Camino Bajo del Castillo s/n, 28692 Villanueva de la Cañada, Madrid, Spain
[3] Institut de Recherche en Astrophysique et Planétologie, CNRS, Université de Toulouse, CNES, Toulouse, France
[4] Space Sciences Laboratory, University of California, Berkeley, CA 94720-7450, USA
[5] Physics Department, University of California, Berkeley, CA 94720-7300, USA
[6] Harvard-Smithsonian Center for Astrophysics, 60 Garden Street, Cambridge, MA, 02138, USA
[7] LIRA, Observatoire de Paris, Université PSL, Sorbonne Université, Université Paris Cité, CY Cergy Paris Université, 92190 Meudon, France
[8] Swedish Institute of Space Physics, Uppsala, Sweden
[9] Department of Physics and Astronomy, Uppsala University, Uppsala, Sweden
[10] Istituto per la Scienza e la Tecnologia dei Plasmi, Consiglio Nazionale delle Ricerche, 70126 Bari, Italy

## Abstract

We investigate a class of ion-scale magnetic solitary structures in the solar wind, characterized by distinct magnetic field enhancements and bipolar rotations over spatial scales of several proton inertial lengths. These structures are revisited using high-resolution data from the Solar Orbiter and Parker Solar Probe missions. Using a machine learning–based method, we identified nearly a thousand such structures, providing new insights into their evolution and physical properties. Statistical analysis shows that these structures are more abundant closer to the Sun, with occurrence rates peaking around 30–40 $R_\odot$ and decreasing farther out. High-cadence measurements reveal that these structures are predominantly found in low-beta ($\beta \leqslant 1$) environments, with consistent fluctuations in density, velocity, and magnetic field. Magnetic field enhancements are often accompanied by plasma density drops, which, under near-pressure balance, limit field increases. This leads to small fractional field enhancements near the Sun (approximately 0.01 at 20 $R_\odot$), making detection challenging. Magnetic field variance analysis indicates that these structures are primarily oblique to the local magnetic field. Alfvénic velocity–magnetic field correlations suggest that most of these structures, unlike most near-Sun solar wind fluctuations, exhibit sunward-directed Alfvénic polarization in the plasma frame. We compare these findings with previous studies, discussing possible generation mechanisms and their implications for the turbulent cascade in the near-Sun Alfvénic solar wind. While these structures might be Alfvénic solitons, vortices, or flux ropes, we refrain from a definitive classification pending further evidence. Further high-resolution observations and simulations are needed to fully understand their origins and impacts.

## 1. Introduction

The solar wind, a continuous supersonic plasma flow from the Sun, serves as an ideal natural laboratory for studying turbulence in collisionless plasmas (R. Bruno & V. Carbone 2013). Despite decades of research, fundamental questions remain about the nature of turbulent fluctuations across the cascade and the mechanisms driving energy dissipation in the solar wind. The solar wind exhibits turbulence characterized by fluctuations spanning a broad range of spatial and temporal scales (e.g., R. Bruno & V. Carbone 2013; D. Verscharen et al. 2019)—where energy cascades from large inertial scales, characterized by power-law behavior similar to fluid turbulence (A. Kolmogorov 1941), with most large-scale fluctuations in the fast solar wind often dominated by highly Alfvénic, outward-propagating plasma (e.g., J. W. Belcher & L. Davis 1971; R. Bruno & V. Carbone 2013)—to ion and electron kinetic scales, where spectral steepening and kinetic effects emerge. At these smaller kinetic scales, plasma behavior becomes increasingly influenced by ion and electron dynamics, altering the nature of the fluctuations (e.g., W. H. Matthaeus & M. Velli 2011; O. Alexandrova et al. 2013).

Although the A. Kolmogorov (1941) framework assumes self-similar scale-invariant fluctuations, real turbulence, such as that observed in the solar wind, exhibits intermittency with localized nonuniform energy concentrations that intensify toward smaller scales. This intermittency leads to deviations from Gaussian statistics and often manifests as coherent structures, which may localize energy and contribute significantly to dissipation (D. Biskamp 1993; U. Frisch 1995). These structures are not merely fluctuations but distinct phenomena associated with specific physical processes, providing valuable insight into the interplay between turbulence and kinetic effects. Particularly at ion scales, such as the ion inertial length ($d_i = c/\omega_{pi}$, where $\omega_{pi} = q_i \sqrt{n_i/(m_i \epsilon_0)}$) and the ion Larmor radius ($\rho_i = V_\perp m_i/(q_i B_0)$, with $V_\perp = \sqrt{k_B T_\perp / m_i}$), these structures could contribute to plasma heating and particle acceleration and are critical for understanding localized energy transfer and dissipation processes in the solar wind (R. J. Leamon et al. 1998; S. D. Bale et al. 2005; C. W. Smith et al. 2006; O. Alexandrova et al. 2013).

In situ observations have identified a variety of ion-scale structures in the solar wind, including current sheets (e.g., C. Cattani 2004; K. T. Osman et al. 2011, 2012; S. Perri et al. 2012; A. Greco & S. Perri 2014), Alfvén vortices (e.g., S. Lion et al. 2016; D. Perrone et al. 2016, 2017; O. W. Roberts et al. 2016), and magnetic holes or dark solitons (e.g., M. Ryutova & H. Hagenaar 2007; D. Perrone et al. 2016). These structures can arise from a variety of processes, including kinetic instabilities (e.g., mirror modes), nonlinear wave steepening, and turbulence-driven relaxation. For instance, quasi-static compressive structures, such as magnetic holes and mirror-mode structures, arise from the mirror-mode instability in high-$\beta$ plasmas under conditions of $T_{\perp} > T_{\parallel}$ (D. Winterhalter et al. 1994; T. L. Zhang et al. 2009; S. Yao et al. 2013). On the other hand, Alfvén vortices can emerge through turbulence-driven relaxation processes that minimize energy by aligning current and vorticity (e.g., M. Imbrogno et al. 2024).

In this study, we focus on a distinct class of ion-scale structures manifesting as localized, nonlinear wave packets. First identified in Ulysses data as rare events (A. Rees et al. 2006), these structures are characterized by clear magnetic field enhancements and bipolar rotations in one or two field components, with a scale size of approximately 30 $d_p$. They exhibit distinctive banana-shaped magnetic hodograms identified through minimum-variance analysis (MVA; B. U. Ö. Sonnerup & M. Scheible 1998). Kinetic simulations by K. Baumgärtel et al. (2007) suggest that kinetic solitary structures, resembling obliquely propagating Alfvén wave pulses with quasi-circular or banana-type polarization, could account for several observational features reported by A. Rees et al. (2006). The former analysis, using Ulysses data, however, was limited by the temporal resolution of the available plasma instruments, restricting detailed plasma studies within these structures. With recent advancements, high-resolution data from the Parker Solar Probe (PSP) and Solar Orbiter (SO) missions now enable the identification of new candidates across varying heliocentric distances, as illustrated by the examples in Figure 1. With data from SO and PSP, we can now study the evolution of these structures and reveal unprecedented plasma signatures within them, including proton density, velocity, temperature, and electron density from onboard instruments. We conducted an extensive search through PSP and SO magnetic field data, covering 5 yr (2019–2023), with a particular focus on PSP encounters within 0.25 au. This effort has led to the identification of nearly a thousand such structures, enabling statistical analysis to address key questions: What are the physical characteristics of these solitary structures? How do they form and evolve within the solar wind? How do they contribute to the broader solar wind turbulence?

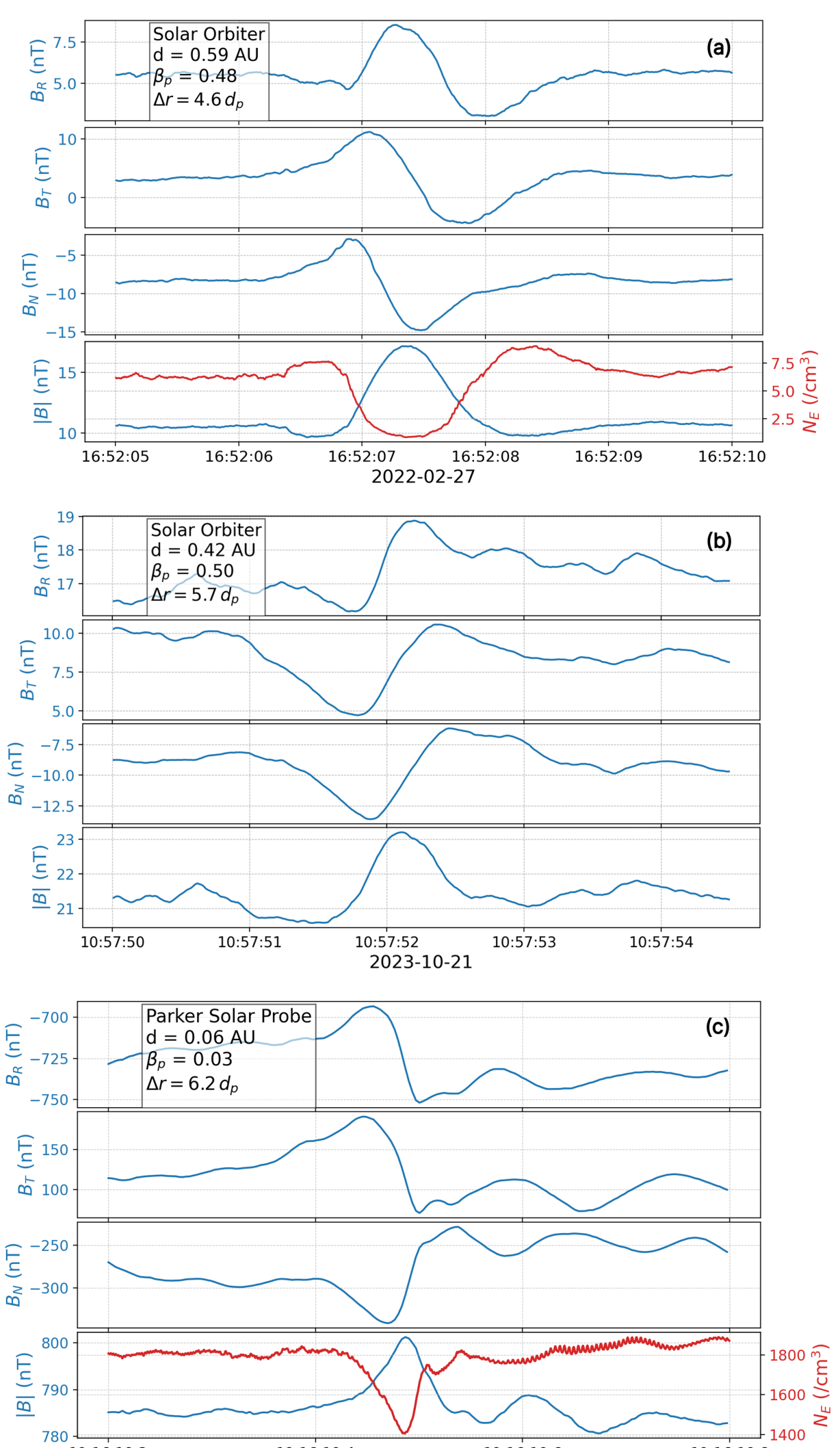

**Figure 1.** Examples of ion-scale solitary structures observed by SO (a), (b) and PSP (c), showing magnetic field profiles and plasma density in the RTN frame. Each panel presents $B_R$, $B_T$, $B_N$, $|B|$, and high-resolution electron number density $N_E$ during the event. The legend specifies the heliocentric distance to the Sun in astronomical units, the proton beta ($\beta_p$), and the event width normalized by the proton inertial length ($d_p$) for each event. Note: $N_E$ is not available for event (b).

## 2. Data and Methods

The data used in this study for the identification and analysis of solitary structures in the solar wind were sourced from SO data and included magnetic field measurements from the magnetometer (T. S. Horbury et al. 2020), proton velocity, density, and temperature from the Solar Wind Analyzer—Proton Alpha Sensor (C. J. Owen et al. 2020), and high-resolution electron density from the Radio and Plasma Waves (M. Maksimovic et al. 2020; Y. V. Khotyaintsev et al. 2021) instrument. PSP data included magnetic field measurements and electron density from the FIELDS suite (S. D. Bale et al. 2016), along with proton velocity, density, and temperature from the Solar Wind Electrons Alphas and Protons—Solar Probe Analyzers instrument (J. C. Kasper et al. 2016).

Traditional methods for identifying ion-scale structures, such as wavelet transforms followed with non-Gaussianity analysis, (e.g., D. Perrone et al. 2016, 2017) and partial variance of increments (e.g., A. Greco et al. 2018), are effective but require substantial manual effort when analyzing extensive solar wind data sets from modern missions. This challenge is particularly pronounced when reviewing data sets from the inner heliosphere, where turbulence further complicates detection. To complement existing detection techniques, we developed a supervised machine learning (ML) classifier to automate the search across years of high-resolution magnetic field data.

We constructed a training data set using traditional intermittency detection methods that focus on identifying the most significant upward fluctuations ("blips"). Specifically, an initial visual inspection of high-resolution magnetic field data identified candidate events with durations of approximately 0.5 s for PSP and 1.0 s for SO. Based on these characteristic durations, we applied continuous wavelet transforms to the full time series of fractional magnetic field fluctuations ($\delta|B|/\langle|B|\rangle$) using a Mexican-Hat wavelet function. We then extracted the wavelet coefficients corresponding to the target scale and applied a stringent 99.99th percentile threshold to isolate the most intermittent time stamps for each day's magnetic field data. However, these scales did not constrain the actual durations of the selected events during subsequent manual inspection using the original magnetic field data, which incorporated MVA and hodogram creation (B. U. Ö. Sonnerup & M. Scheible 1998). Key selection criteria included (1) significant magnetic field enhancement, (2) a clear bipolar rotation in at least one RTN and MVA component with associated variations in others, and (3) well-defined hodogram patterns resembling banana/arc or quasi-circular shapes. Through this process, 466 manually verified events (157 from SO and 309 from PSP) were identified, forming the basis for the training data set of our ML classifiers. We subsequently applied data augmentation techniques to expand this training set. Several ML algorithms were tested, with the random forest model (L. Breiman 2001) demonstrating the highest precision, interpretability, and effectiveness in identifying new events within unseen candidate time windows generated using the same wavelet transform and threshold strategy. To prioritize true positives, we employed a conservatively high classification threshold (0.9), accepting some missed detections (false negatives) in favor of greater overall accuracy.

This ML-based identification method led to the discovery of 974 events, significantly enhancing detection efficiency. The final event data set comprises structures with temporal durations ranging from approximately 0.1 to 1.0 s, determined through Gaussian fitting of the magnetic field enhancement profiles, with the extracted FWHM serving as a measure of event duration. Applying Taylor's hypothesis and the local solar wind speed, these durations correspond to spatial scales of approximately 5–200 km for PSP events and 100–600 km for SO events. Detailed statistical analyses of their profiles and characteristics, occurrence and evolution, Alfvénic signatures, and comparisons with nonevent periods are discussed in Section 3. Since the primary focus of this Letter is on the physical interpretation and implications of the results, the details of the ML framework are presented separately in Y. Yang (2024).

## 3. Results

### 3.1. Profiles and Characteristics

The event illustrated in Figure 1(a) represents a typical, well-defined example observed by SO, lasting approximately 1 s. Its smooth magnetic profile features a pronounced field enhancement resembling a Gaussian blip with two dips on either side, closely matching the shape of a Mexican-Hat wavelet function. This enhancement is accompanied by distinct spikes in all three RTN components, with a return to the initial values, suggesting a solitary structure rather than a boundary crossing. These rotations persist longer than the magnetic field enhancement itself. A banana-shaped hodogram, observed in the $B_{\mathrm{int}}$ versus $B_{\max}$ plot within the MVA frame, is similar to the features reported by A. Rees et al. (2006), though not explicitly shown here. Additionally, plasma measurements of proton and electron densities reveal that the magnetic field enhancement is often anticorrelated with a plasma density dip, as shown in the fourth panel of the figure. Correlated electric field fluctuations are also found to accompany these magnetic variations, but these are challenging to interpret given instrumental and spacecraft effects, so we do not discuss them in detail in this work. While this example is clear, the final event data set includes less distinct cases where magnetic profiles are less smooth, and rotations in field components are incomplete or less pronounced relative to background fluctuations. Such less distinct magnetic rotations observed in some events may arise from more complex or noncylindrical topologies: a single-spacecraft trajectory can sample only part of the structure, yielding partial rotations in the measured field components. In addition, the turbulent background adds complexity and challenges in consistently applying strict classification criteria for these structures.

Although these events can be isolated, a substantial proportion were found to cluster. Specifically, 94% of events in the PSP data set and 54% in the SO data set occurred within 24 hr of another event, and 57% of events in PSP and 38% in SO reoccurred within an hour. This suggests that certain conditions may favor the formation of multiple events. Such multievent clusters are primarily observed in regions where the background magnetic field is both low and stable, exemplified in Figure 2. The event shown in Figure 1(a) falls within this period. In panels (b) and (c) of Figure 2, we observe consistent patterns of associated variations among magnetic field profiles and density profiles for this cluster, resembling the single event example. In panel (d), the prominent magnetic field enhancement, coupled with a modest plasma density drop, results in a more significant increase in magnetic pressure than the decrease in plasma pressure (calculated using higher-resolution electron density as a proxy for proton density and proton temperature data). This suggests that these structures are magnetically dominated, perhaps exhibiting a temporary slight pressure imbalance that leads to an increase in overall pressure within the event and a reduction in plasma $\beta$, as shown in panel (e) of Figure 2; $\beta$ remained below 1 throughout this period.

### 3.2. Radial Evolution

Figure 1 presents structures observed at varying distances from the Sun: SO examples at 0.6 au (panel (a)) and 0.4 au (panel (b)), and a closer PSP example at 0.06 au (panel (c)). Consistent correlated fluctuations in the magnetic field, density, and electric field are observed across these examples. While not all events are as clear as these, all exhibit short and smooth magnetic field enhancement with associated bipolar rotations in the components. Notably, the fractional field enhancements typically decrease as heliocentric distance decreases, from $\delta|B|/\langle|B|\rangle \sim 0.75$ in Figure 1 example (a), $\sim$0.09 in example (b), to $\sim$0.03 in example (c). This pattern suggests that the maximum amplitude of the magnetic blips is constrained by plasma density drops under the assumption of total-pressure conservation. In other words, if we assume that pressure balance exists inside and outside the structures,

$$P_{\text{outside}} = P_B + P_P = P_{\text{inside}} = P_B' + P_P'. \quad (1)$$

Then, with the plasma density approaching zero, the minimum plasma pressure $P_P'$ inside the blip is nearly zero, and the maximum magnetic pressure $P_B'$ inside the blip is bounded by

$$P_B' \leqslant (1 + \beta) P_B. \quad (2)$$

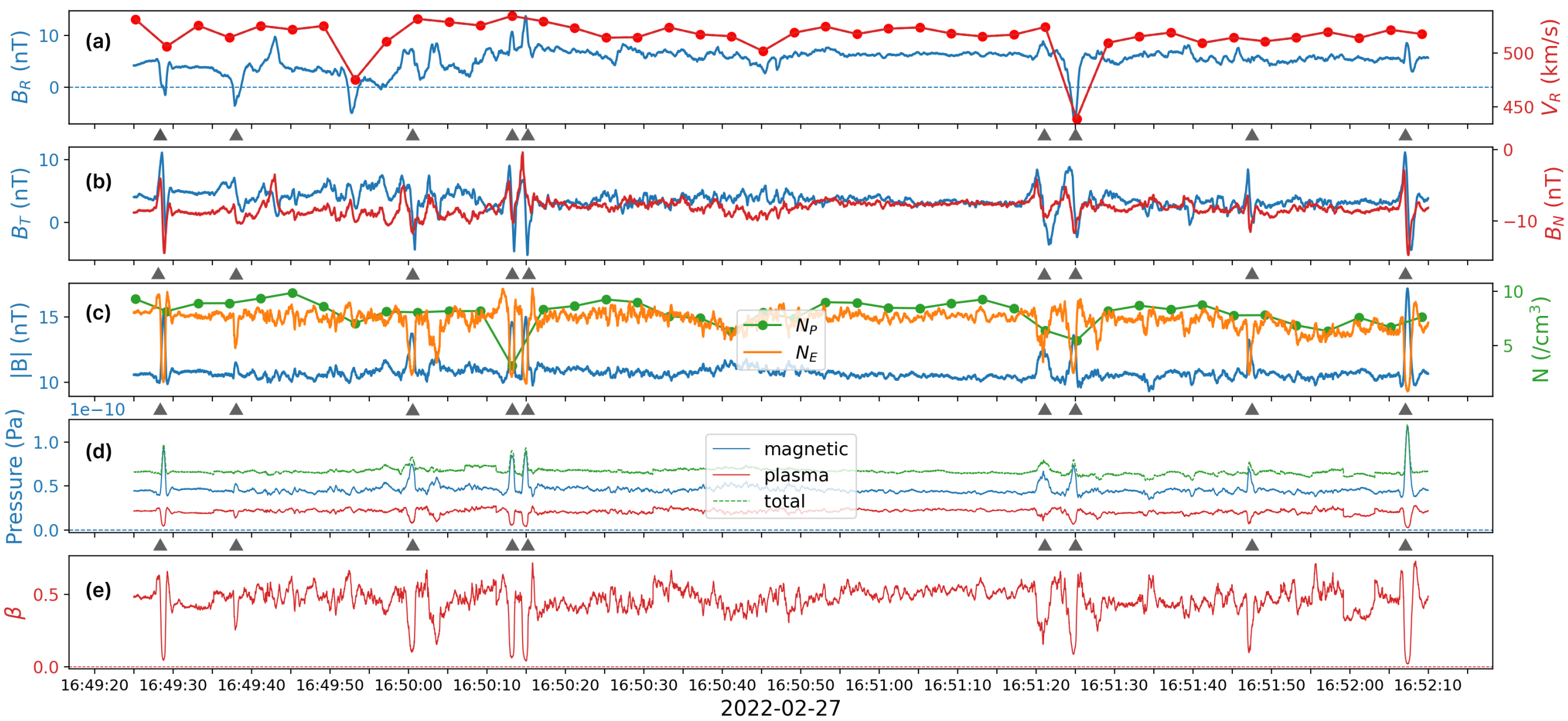

**Figure 2.** An example of a multiple-event period observed by SO. Panels show (a) $B_R$ and $V_R$, (b) $B_T$ and $B_N$; (c) $|B|$, $N_p$ (proton number density), and $N_E$ (electron number density); (d) magnetic, plasma (evaluated using high-resolution electron number density and proton temperature), and total pressures; and (e) plasma $\beta$, calculated using the magnetic and plasma pressures. Triangles across the panel grid lines indicate the occurrence of each event.

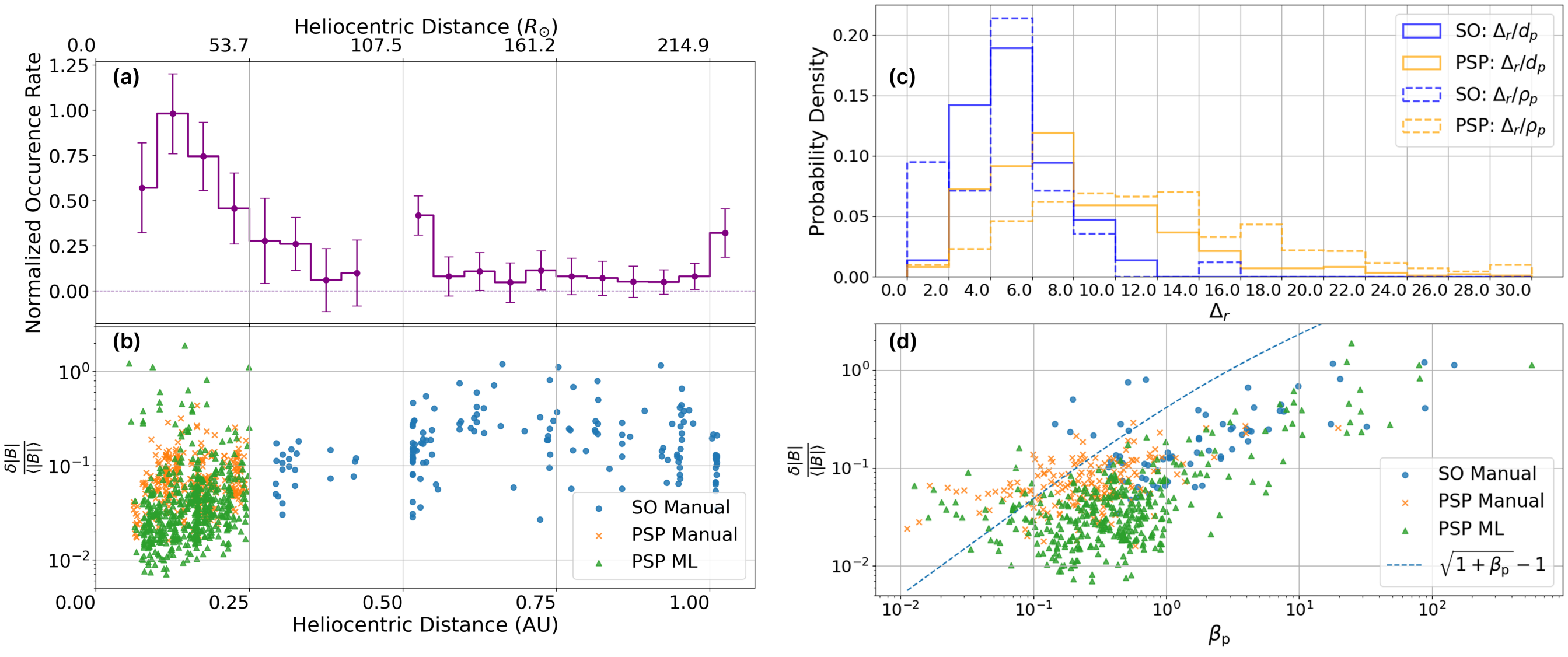

**Figure 3.** Statistical analysis results of the ion-scale solitary structures data set. (a) Normalized occurrence rate (events per day), representing the aggregated occurrence rate across data sets adjusted for data availability and scaled to address methodological differences and (b) fractional magnetic field enhancement ($\delta|B|/\langle|B|\rangle$), both presented as functions of radial distance. Error bars in (a) represent standard errors, assuming occurrences are Poisson random variables. (c) Distributions of convected event radial widths, $\Delta_r$, normalized by proton inertial length ($d_p$) and proton Larmor radius ($\rho_p$), with separate frequency plots for events from Solar Orbiter (SO) and Parker Solar Probe (PSP). (d) $\delta|B|/\langle|B|\rangle$ plotted against proton beta ($\beta_\mathrm{p}$) for each event. The blue dashed line represents the theoretical upper bound $\sqrt{1+\beta_\mathrm{p}}-1$ on $\delta|B|/\langle|B|\rangle$, restricted by $\beta_\mathrm{p}$ under pressure-balance conditions. Legends label data sources by method: manually identified events (SO manual: blue circles; PSP manual: orange crosses) and ML-detected events (PSP ML: green triangles).

Hence, the change in magnetic field strength is limited by

$$\frac{|B'|}{|B|} = \sqrt{\frac{P_B'}{P_B}} \leqslant \sqrt{1+\beta}. \tag{3}$$

And the fractional change in magnetic field strength is expressed as

$$\frac{\delta|B|}{|B|} = \frac{|B'| - |B|}{|B|} \leqslant \sqrt{1+\beta} - 1. \tag{4}$$

It is important to note that this pressure-balance condition is used here solely as an upper bound for the fractional field enhancement; it is not intended to assert that all events are perfectly pressure balanced or strictly standing. In fact, the dashed line in Figure 3 merely delineates the maximum $\delta|B|$ expected under idealized total-pressure conservation, and our events generally lie below this limit, which helps explain why $\delta|B|/\langle|B|\rangle$ decreases with decreasing heliocentric distance (and lower plasma $\beta$), a trend also shown in panel (b) of Figure 3.

These findings reveal the detection challenges within 0.25 au, where minimal $\delta|B|/\langle|B|\rangle$ underscores the need for ML techniques to identify these structures effectively.

Furthermore, a small fraction of events—3% for SO and 6% for PSP—exceed the upper limit set by Equation (4). Additional comparisons of the magnetic and plasma pressures (e.g., in Figure 2) typically reveal a slight excess in magnetic pressure. These deviations could arise from slight pressure imbalances in the structures or from limitations in beta estimation and measurement uncertainties. We note that our plasma pressure is estimated using proton density and temperature alone, which may underestimate the total pressure by neglecting contributions from electrons and other ions. Nonetheless, these results support the view that the observed $\delta|B|/\langle|B|\rangle$ is probably constrained by pressure balance, even if perfect equilibrium is not achieved.

Although detecting these structures is more challenging closer to the Sun due to lower $\delta|B|/\langle|B|\rangle$, our analysis of occurrence rates as a function of heliocentric distance shows they are more abundant nearer to the Sun. These rates represent aggregated event counts per day across data sets, normalized for data availability and scaled to account for methodological differences. The scaling process involved comparing occurrence rates at overlapping distances between ML and manual detection methods, as well as between PSP and SO data sets, accounting for variations in wavelet transform scales. Panel (a) of Figure 3 shows the occurrence rate peaks around 0.1 au and decreases with distance, suggesting a possible decay of these structures. The lower occurrence rates at the smaller distances, within 0.05 au, suggest that these structures may either form beyond this region or become undetectable in the near-Sun's turbulent magnetic field environment with its stronger field.

To investigate the physical scales of these events, we approximated the event width as the radial extent of the convected magnetic field enhancement since the temporal profiles of the RTN magnetic field components remain essentially unchanged within each event. We obtain this scale by multiplying the temporal FWHM, derived from a Gaussian fit to the field magnitude, with the radial relative velocity between the spacecraft and the solar wind. This frozen-in (Taylor) hypothesis is equivalent to slow propagation of the structures in the plasma frame, consistent with the large angles observed between the minimum-variance direction $\boldsymbol{e}_{\min}$ and the background field $B_0$ (Section 3.3). The Gaussian fit also yields the peak amplitude ($\delta|B|$) and the average magnetic field magnitude ($\langle|B|\rangle$). Spatial widths were further normalized by the proton inertial length ($d_p$) and proton Larmor radius ($\rho_p$) for analysis. The width distributions normalized by $d_p$ were more consistent between SO and PSP data sets comparing those normalized by $\rho_p$, exhibiting a clear peak between 2 and 12 $d_p$ as seen in Figure 3(c), broadly consistent with Alfvén solitons reported by A. Rees et al. (2006), as well as current filaments observed in O. Alexandrova et al. (2004) and Alfvén vortices observed by e.g., O. Alexandrova et al. (2006), D. Perrone et al. (2016, 2017), and T. Wang et al. (2019).

It is challenging to use single-spacecraft measurements to definitively distinguish between propagating waves and structures that are merely convected, i.e., 2D cylindrical flux ropes (V. Lanabere et al. 2020) or Alfvén vortices (V. I. Petviashvili & O. A. Pokhotelov 1992; O. Alexandrova et al. 2006; T. Wang et al. 2019). In our width estimation, we found that assuming either a stationary or a propagating structure at the local Alfvén speed $V_A$ does not significantly alter the spatial width probability distribution due to the structures' slow propagation. Future multispacecraft analyses will be necessary to further elucidate their topology.

### 3.3. Orientation, Alfvénic Signatures, and Background Conditions

Magnetic field variance analysis revealed large eigenvalue ratios ($\lambda_{\rm int}/\lambda_{\min}$, $\lambda_{\max}/\lambda_{\rm int}$), indicating a well-defined fluctuation geometry. To investigate the orientation of these structures relative to the background field $B_0$, we compared them to randomly sampled nonevent periods of comparable duration. As shown in Figure 4, events are predominantly observed when $B_0$ is nearly perpendicular to the radial (sampling) direction (panel (a)). In addition, the $\boldsymbol{e}_{\min}$ directions also tend to be nearly perpendicular to $B_0$ (panel (b)). Taken together, these results are consistent with structures whose wavevectors lie at a large angle to $B_0$ and so are therefore preferentially observed when the field is transverse to the flow. Furthermore, in panel (c), the angle between $\boldsymbol{e}_{\max}$ and $B_0$ is typically close to 90°, consistent with the magnetic fluctuations being predominantly perpendicular to $B_0$: even though the magnetic field magnitude varies within these structures, variations in the transverse components are still larger.

As can be seen from Figure 2(a), despite the limited plasma time resolution (4 s), there is clear evidence for velocity variations associated with these events. The positive correlations between $\delta B_R$ and $\delta V_R$ within this interval, with a positive background radial field component, is consistent with an Alfvénic correlation in a sunward sense (J. W. Belcher & L. Davis 1971), in contrast to most fluctuations in the inner solar system. This Alfvénic correlation in a sunward sense is evident for most events, with 640 events from PSP and 74 from SO exhibiting this trend, while antisunward events were comparatively rare. Correlations were evaluated within a time window on the order of the local velocity cadence, ensuring three velocity points per estimate. The large-scale field polarity was defined as the sign of $\langle B_R\rangle$ averaged over the 15 s centered on each event. Limitations in velocity data, particularly in the SO data set, restricted analysis for some events, and the low temporal resolution of velocity data introduced additional uncertainties in confirming polarization direction for certain structures.

Additionally, we compared plasma conditions between event and nonevent periods. Results indicate that events tended to occur at slightly elevated proton temperatures relative to nonevent periods. Both PSP and SO data show consistently higher temperatures within the identified events. While this may suggest that solitary structures preferentially form in regions of elevated thermal energy, it is also possible that these structures are intrinsically dissipative, consistent with their intermittent nature, and contribute to local plasma heating. For ambient solar wind speed, PSP data reveal that events occur at a higher average speed (400 km s$^{-1}$ versus 300 km s$^{-1}$) compared to nonevents, whereas SO data show no significant speed separation, with both events and nonevents averaging around 400 km s$^{-1}$. This suggests a possible link between faster wind speeds and event occurrence. Plasma beta distribution shows no significant difference between events and nonevents, both typically lying within regions with $\beta \leqslant 1$. Temperature anisotropy analysis also reveals no strong association of these structures with proximity to any specific plasma instabilities, such as mirror or fire hose, indicating these structures typically form under stable solar wind conditions without a distinct tendency toward these instabilities.

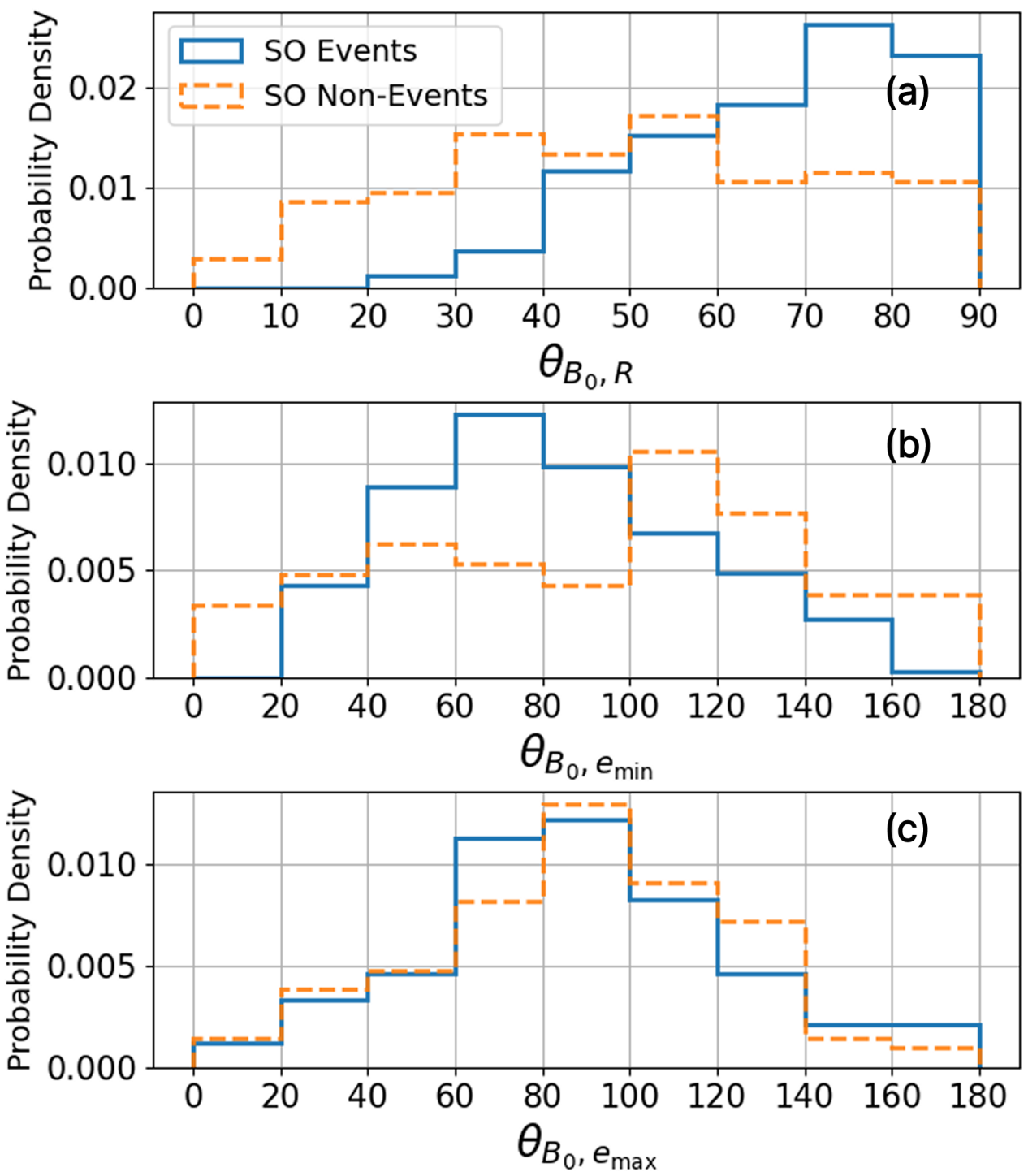

**Figure 4.** Normalized distributions of angle $\theta$ between the background magnetic field ($B_0$) and (a) the radial direction, (b) the minimum-variance direction ($\boldsymbol{e}_{min}$), and (c) the maximum variance direction ($\boldsymbol{e}_{max}$). Legends are shared across all panels and represent events from Solar Orbiter (SO events) and randomly sampled intervals without events (SO nonevents).

## 4. Discussion

The analysis presented here demonstrates that the isolated structures first identified by A. Rees et al. (2006) beyond 1 au are significantly more common closer to the Sun (0.1 au) but are challenging to detect due to their relatively smaller amplitudes. These structures exhibit well-defined magnetic field profiles with distinct field magnitude enhancements accompanied by sharp decreases in plasma density to very low values. With scales of a few proton inertial lengths, their minimum-variance directions are primarily oblique to the background magnetic field, and around these structures, Alfvénic $V$–$B$ correlations are opposite to most plasma in the inner heliosphere, consistent with a sunward polarization sense. Variations across the three magnetic field components, indicating both field rotation and compression, distinguish these structures from other purely compressive phenomena, such as magnetic holes or mirror modes, which typically show linear polarization with variations confined to a single direction.

The observed behavior could be related to Alfvénic solitons (E. Mjølhus 1976; S. R. Spangler & J. P. Sheerin 1982), localized MHD structures driven by transverse magnetic field oscillations that exhibit pronounced variations in $|B|$. However, under ion-scale dispersive effects, such solitons and wave packets may become unstable to parametric modulational instability, leading to steepening and enhancement of $|B|$ perturbations (B. Buti et al. 2000), also causing the structures' disruptions. This steepening can generate parallel electric fields that efficiently accelerate ions, potentially forming proton beams (e.g., L. Matteini et al. 2010) and significantly affecting ion kinetics in space plasmas. Opposite Alfvénic $V$–$B$ correlations relative to the background Alfvénic fluctuations are observed around the structures, indicating a sunward-directed polarization sense, which suggests parametric decay instability (N. F. Derby 1978; M. L. Goldstein 1978) as a possible generation mechanism. Under quasi-parallel propagation, Alfvénic solitons typically exhibit positive correlations between magnetic field magnitude and density fluctuations for $\beta < 1$ (B. Buti et al. 2000), whereas for sufficiently oblique propagation angles, they can maintain anticorrelated $\delta n$–$\delta B$ fluctuations, even at $\beta \sim 1$, suggesting pressure balance (K. Baumgärtel et al. 2007), and such anticorrelation is also discussed in a recent study (A. Mallet 2023), which derived weakly nonlinear evolution equation for large-amplitude, very oblique Alfvén waves with length scales comparable to ion inertial lengths under low-$\beta$ conditions. The features of the structures, particularly their large $k_\perp$ component (see Section 3.3), then suggest a link to these simulation results.

Meanwhile, this significant $k_\perp$ component suggests a potential connection to Alfvén vortices, which have been extensively studied and observed across a wide range of scales and plasma environments. They have been identified in both slow and fast solar winds (e.g., D. Perrone et al. 2016, 2017), with anticorrelated field-density fluctuations observed by T. Wang et al. (2019), as well as closer to the Sun (A. Vinogradov et al. 2024). Early observations in the Earth's magnetosheath found Alfvén vortices at scales of approximately 10 $d_p$ (O. Alexandrova et al. 2004), and similar structures were later reported in the solar wind at comparable scales (S. Lion et al. 2016). Theoretical models describe these structures across multiple regimes, from MHD descriptions of large-scale vortices (V. I. Petviashvili & O. A. Pokhotelov 1992; O. Alexandrova et al. 2006; O. Alexandrova & J. Saur 2008) to fluid models for moderate-$\beta$ plasmas that account for compressible effects at ion scales (D. Jovanović et al. 2020).

While our events share some similarities with the above discussed structures, they also exhibit distinct features that suggest they may represent a special case rather than conventional ones. First, we deliberately focus on events with large $\delta|B|/|B|$ that exhibit bipolar rotations against a smooth background. Second, around our events, there is an an Alfvénic $V$–$B$ correlation in the sunward direction, which contrasts with the dominant outward-propagating Alfvénic streams of the solar wind. Consequently, it remains unclear whether these events belong to the same category as previously identified structures or form a distinct subset. Further analysis is needed to determine their precise nature.

The clustering of these events near the Sun and their distinct properties may be linked to the large-amplitude antisunward propagating fluctuations, dominated by switchbacks in the inner heliosphere, as observed by S. D. Bale et al. (2019) and J. C. Kasper et al. (2019). Parametric decay of these fluctuations (e.g., L. Matteini et al. 2010; M. Shi et al. 2017; T. A. Bowen et al. 2018) could generate compressive sunward-propagating structures, consistent with those observed here. The peak occurrence rate near $\sim 30 R_S$, as suggested by Figure 3(a), might correspond to the growth and saturation of $\delta B/B \approx 1$ switchbacks around this distance (Z. Huang et al. 2023). These events may occur more frequently than switchbacks closer to the Sun but become less frequent farther out. However, no definitive conclusions can be drawn, leaving a direct investigation of the correlation between their occurrence rates as future work. The apparent decay of these structures beyond 0.1 au may result from interactions with background fluctuations. With opposite polarization sense, these structures likely interact with switchbacks, decaying in the process while potentially triggering energy transfer within the switchbacks themselves, thereby contributing to the turbulent cascade. Such events might represent transient mediators of the decay process, contributing to the conversion of switchbacks into a broadband turbulent cascade (T. S. Horbury et al. 2023) that ultimately heats and drives the solar wind (N. E. Raouafi et al. 2023).

In summary, this study reports statistical results of a distinct type of ion-scale Alfvénic solitary structures in the solar wind, observed by SO and PSP. However, ambiguity remains regarding their precise classification and origin, necessitating further theoretical and observational work to explore the relationship between oblique Alfvénic solitons, Alfvén wave packets, and Alfvén vortices at ion scales. Their prevalence at smaller heliocentric distances and their relation with switchbacks deserve further investigation. Most importantly, the apparent sunward polarization of these structures and their abundance in the solar wind suggest potential interactions with the dominant outward-propagating plasma flows, indicating mechanisms for energy transfer or dissipation. This necessitates further work to fully understand their generation mechanisms and their roles in solar wind turbulence.

## Acknowledgments

Solar Orbiter magnetometer operations are funded by the UK Space Agency (grant ST/X002098/1). Y.Y. and T.S.H. were supported by STFC grant ST/W001071/1. J.H.W. is supported by the STFC under studentship ST/X508433/1. The RPW instrument has been designed and funded by CNES, CNRS, the Paris Observatory, the Swedish National Space Agency, ESA-PRODEX and all the participating institutes. The SWA/PAS instrument was funded by CNRS and CNES. AL acknowledges the support of the PRIN 2022 project ``2022KL38BK — The ULtimate fate of TuRbulence from space to laboratory plAsmas (ULTRA)'' (Master CUP B53D23004850006) by the Italian Ministry of University and Research, funded under the National Recovery and Resilience Plan (NRRP), Mission 4 – Component C2 – Investment 1.1, ``Fondo per il Programma Nazionale di Ricerca e Progetti di Rilevante Interesse Nazionale (PRIN 2022)'' (PE9) by the European Union – NextGenerationEU. Y.K. is supported by the Swedish National Space Agency grant 2024-00108. M.L.S., D.E.L., and R.L. are supported by the Parker Solar Probe mission SWEAP investigation, under NASA contract NNN06AA01C.

## Use of Libraries and Tools

This study utilized open-source libraries and scientific computing tools, including `NumPy` (C. R. Harris et al. 2020), `pandas` (J. Reback et al. 2020), `SciPy` (P. Virtanen et al. 2020), `Matplotlib` (J. D. Hunter 2007), `SunPy` (The SunPy Community et al. 2020), `scikit-learn` (F. Pedregosa et al. 2011), and `TensorFlow` (M. Abadi et al. 2016).

## ORCID iDs

Yufei Yang (杨宇菲) https://orcid.org/0009-0004-8349-3281
Timothy S. Horbury https://orcid.org/0000-0002-7572-4690
Domenico Trotta https://orcid.org/0000-0002-0608-8897
Lorenzo Matteini https://orcid.org/0000-0002-6276-7771
Joseph H. Wang https://orcid.org/0009-0008-4095-9175
Andrey Fedorov https://orcid.org/0000-0002-9975-0148
Philippe Louarn https://orcid.org/0000-0003-2783-0808
Stuart D. Bale https://orcid.org/0000-0002-1989-3596
Marc Pulupa https://orcid.org/0000-0002-1573-7457
Davin E. Larson https://orcid.org/0000-0001-5030-6030
Roberto Livi https://orcid.org/0000-0002-0396-0547
Michael L. Stevens https://orcid.org/0000-0002-7728-0085
Milan Maksimovic https://orcid.org/0000-0001-6172-5062
Yuri V. Khotyaintsev https://orcid.org/0000-0001-5550-3113
Andrea Larosa https://orcid.org/0000-0002-7653-9147